# Systemic Consequences of Disorder in Magnetically Self-Organized Topological MnBi$_2$Te$_4$/(Bi$_2$Te$_3$)$_n$ Superlattices

Joanna Sitnicka,[1] Kyungwha Park,[2] Paweł Skupiński,[3] Krzysztof Grasza,[3] Anna Reszka,[3] Kamil Sobczak,[4] Jolanta Borysiuk,[1] Zbigniew Adamus,[3] Mateusz Tokarczyk,[1] Andrei Avdonin,[3] Irina Fedorchenko,[5] Irina Abaloszewa,[3] Sylwia Turczyniak-Surdacka,[4] Natalia Olszowska,[6] Jacek Kołodziej,[6,7] Bogdan J. Kowalski,[3] Haiming Deng,[8] Marcin Konczykowski,[9] Lia Krusin-Elbaum,[8] and Agnieszka Wołoś[1*]

[1]*Faculty of Physics, University of Warsaw, ul. Pasteura 5, 02-093 Warsaw, Poland*

[2]*Department of Physics, Virginia Tech, 850 West Campus Drive, Blacksburg, VA 24061, USA*

[3]*Institute of Physics, Polish Academy of Sciences, Aleja Lotników 32/46, PL-02668 Warsaw, Poland*

[4]*Faculty of Chemistry, Biological and Chemical Research Centre, University of Warsaw, ul. Zwirki i Wigury 101, 02-089 Warsaw, Poland*

[5]*Kurnakov Institute of General and Inorganic Chemistry, Russian Academy of Sciences, Leninskii prosp. 31, 117901 Moscow, Russia*

[6]*National Synchrotron Radiation Centre SOLARIS, Jagiellonian University, ul. Czerwone Maki 98, 30-392 Cracow, Poland*

[7]*Faculty of Physics, Astronomy and Applied Computer Science, Jagiellonian University, ul. prof. Stanisława Łojasiewicza 11, 30-348 Cracow, Poland*

[8]*Department of Physics, The City College of New York–CUNY, 85 St. Nicholas Terrace, New York, NY 10027, USA*

[9]*Laboratoire des Solides Irradiés, CEA/DRF/lRAMIS, Ecole Polytechnique, CNRS, Institut Polytechnique de Paris, F-91128 Palaiseau, France*

**Abstract**

MnBi$_2$Te$_4$/(Bi$_2$Te$_3$)$_n$ materials system has recently generated strong interest as a natural platform for realization of the quantum anomalous Hall (QAH) state. The system is magnetically much better ordered than substitutionally doped materials, however, the detrimental effects of certain disorders are becoming increasingly acknowledged. Here, from compiling structural, compositional, and magnetic metrics of disorder in ferromagnetic MnBi$_2$Te$_4$/(Bi$_2$Te$_3$)$_n$ it is found that migration of Mn between MnBi$_2$Te$_4$ septuple layers (SLs) and otherwise non-magnetic Bi$_2$Te$_3$ quintuple layers (QLs) has systemic consequences — it induces ferromagnetic coupling of Mn-depleted SLs with Mn-doped QLs, seen in ferromagnetic resonance as an acoustic and optical resonance mode of the two coupled spin subsystems. Even for a large SL separation (n $\gtrsim$ 4 QLs) the structure cannot be considered as a stack of uncoupled two-dimensional layers. Angle-resolved photoemission spectroscopy and density functional theory studies show that Mn disorder within an SL causes delocalization of electron wave functions and a change of the surface band structure as compared to the ideal MnBi$_2$Te$_4$/(Bi$_2$Te$_3$)$_n$. These findings highlight the critical importance of inter- and intra-SL disorder towards achieving new QAH platforms as well as exploring novel axion physics in intrinsic topological magnets.

*Corresponding author: agnieszka.wolos@fuw.edu.pl

## 1. Introduction

Searching for materials appropriate for realization of the quantum anomalous Hall (QAH) effect or axion insulator state [1–6] led recently to an explosive interest in MnBi$_2$Te$_4$ family, and in particular in a self-organized MnBi$_2$Te$_4$/(Bi$_2$Te$_3$)$_n$ with the magnetic MnBi$_2$Te$_4$ septuple (SLs) layers structurally and compositionally compatible with the non-magnetic Bi$_2$Te$_3$ topological insulator.[7–13] When MnBi$_2$Te$_4$ is located on the top surface of the structure the topological surface states are expected to appear into the magnetic material, which in contrast to weak proximity effects studied earlier [14] causes giant opening of the Dirac gap and a strong modification of spin texture of the topological surface states.[7,11,15–17] Manganese self-organizes planarly while also doping Bi$_2$Te$_3$, where at low doping level it preferentially substitutes on the Bi sites, akin to doping in Bi$_2$Se$_3$.[18] Increased doping results in Mn entering an interstitial position in the van der Waals gap.[19] When doping exceeds about 2 at.%, Mn self-organizes into MnBi$_2$Te$_4$ septuple layers within the Bi$_2$Te$_3$ matrix.[3,11] Although this system is magnetically much better organized than the substitutionally doped materials,[20] it is not free from the disorder, which can arise from (i) statistical distribution of septuple layers, (ii) magnetic disorder within septuple layers, or (iii) doping of Mn into Bi$_2$Te$_3$ quintuple layers (QLs). These three highlighted effects strongly impact both magnetism and the surface band structure. Thus, in order to get a controllable access to the magneto-topological phenomena the disorder should be understood and mastered.

The disorder effects have been very recently recognized in the family of MnBi$_2$Te$_4$/(Bi$_2$Te$_3$)$_n$ or in compounds containing Sb as a driving force for ferromagnetism (FM) in otherwise antiferromagnetic (AFM) material.[17,21] This led to some basic questions regarding the role of the disorder in a broader sense. In particular, the fundamental question concerns the effects of disorder on surface electronic band structure and on magnetic order. Presently, there is insufficient knowledge of the magnetic ordering of manganese ions in a composite MnBi$_2$Te$_4$/(Bi$_2$Te$_3$)$_n$, complicated by the Mn site mixing, since both intra-layer and inter-layer coupling comes into play, dependent on the Mn-Mn distance in SLs and on the non-uniformly distributed other possible Mn defect sites. The effect of disorder on the surface electronic band structure also requires deeper understanding, due to the abundance of experimentally observed electronic bands and the still somewhat unclear situation with the temperature dependence of the Dirac mass gap.[9,11,17] Recent reports provide disturbing evidence on the fatal influence of disorder on the band inversion necessary to obtain topological insulator phase.[13,21] On the other hand, classifying and controlling the effects of disorder can lead to a better understanding of the QAH state.[22]

In this work, we show results of structural and chemical composition investigations of ferromagnetic MnBi$_2$Te$_4$/(Bi$_2$Te$_3$)$_n$, with *n* between 2 and 12 and the Curie temperature, $T_c$, ranging between 6 K and 13 K, and correlate it with magnetic properties studied by ferromagnetic resonance (FMR) and with properties of electronic surface band structure investigated by angle-resolved photoemission spectroscopy (ARPES) and density functional theory (DFT). Our structural and chemical composition studies show the omnipresent migration of Mn between MnBi$_2$Te$_4$ SLs and Bi$_2$Te$_3$ QLs. In order to investigate ferromagnetism in these disordered structures we applied X-band FMR, a technique which is complementary to standard magnetometric measurements and brings additional information about the magnetic system. In particular, the FMR has been widely applied to ultrathin films allowing investigation of the interlayer exchange coupling.[23–26] The Mn site mixing between SLs and QLs changes this material systemically making it akin to a stack of exchange-coupled ultra-thin magnetic films of Mn$_x$Bi$_{3-x}$Te$_4$ and (Bi,Mn)$_2$Te$_3$. The surface electronic states become delocalized under disorder and the structure of the bands changes compared to the ideal MnBi$_2$Te$_4$/(Bi$_2$Te$_3$)$_n$ as seen both in ARPES and DFT.

**Table 1.** Structural and magnetic parameters of $MnBi_2Te_4/(Bi_2Te_3)_n$: mean distance between SLs ($n$), average Mn concentration in a sample from SEM-EDX and Mn concentration in SLs from TEM-EDX, respectively, critical temperature for bulk FM phase transition ($T_c$) from magnetometric or magnetotransport measurements. Mn concentration in QLs is typically between 0.3 – 1 at. % indicated by TEM-EDX.

| No. | $n$ (QLs) | Mn average (at. %) | Mn in SLs (at. %) | $T_c$ (K) |
|---|---|---|---|---|
| S1 | 12 ± 5 | 2 | 4.4 ± 0.7 | 6 |
| S2 | 4 | 2 | 10.6 ± 1.6 | 13 |
| S3 | 7 ± 1 | 2 | 5.0 ± 0.8 | 10 |
| S4 | 2 | 4 | 7.6 ± 1.1 | 10 |

## 2. Disorder Modifications of Structural Properties and Chemical Composition

All $MnBi_2Te_4/(Bi_2Te_3)_n$ samples studied in this work show almost uniform distribution of the elements in the energy dispersive x-ray analysis (EDX) on submicron scale accessible in scanning electron microscopy (SEM), with the average Mn concentration 2 or 4 at. %, respectively (Table 1). Neither inhomogeneities of the chemical composition nor Mn inclusions are observed. In contrast, transmission electron microscopy (TEM) studies of the same samples performed on nm- or subnanometer scale reveal presence of self-organized SLs of $MnBi_2Te_4$ incorporated into $Bi_2Te_3$ matrix, Figure 1 (a) and (b). Manganese is planarly distributed in the middle of a SL while the SL and the neighboring QL are separated by a van der Waals gap. Although all samples (except for S4) have the same average Mn concentration, Mn can be distributed in different ways. The largest separation between SLs, $n$ = 12 QLs in the $MnBi_2Te_4/(Bi_2Te_3)_n$ formula, is found in S1 sample obtained by the vertical variant of the Bridgman method after the synthesis only. Small statistical distribution with standard deviation (SD) of 5 QLs can be observed, Figure 1 (c). After crystallization, sample S2, SLs are rearranged keeping a preferential distance of 4 QLs and a grouping in a superlattice containing multiples of mostly 3 SLs (can be 3, 6, or more rarely 9 SLs), Figure 1 (f). The SLs extend without breaking their continuity all along the investigated 16 µm-wide lamellae. Sample S3 obtained by the horizontal variant of the Bridgman method has a medium separation between SLs with mean value of about 7 QLs (SD = 1 QL). A break in the continuity of SLs is often observed in this sample where septuple layers interchange with quintuple layers. Finally, sample S4 (horizontal Bridgman) has a very uniform morphology and the smallest distance between SLs of 2 QLs. This sample has higher average Mn concentration, about 4 at.%. Different characteristics of the four samples are summarized in Table 1.

Nano-structural studies show omnipresent migration of Mn between SLs and QLs. EDX analysis performed on nm-scale of TEM reveals a dearth of Mn in SLs and a simultaneous presence of Mn in QLs. The highest Mn concentration in SLs, 10.6 ± 1.6 at.%, is found in sample S2. This agrees within three sigma limit to the nominal value of the $MnBi_2Te_4$ formula (~14 at.%) and simultaneously S2 sample has the highest ferromagnetic phase transition temperature 13 K. The companion sample that did not undergo the second step in Bridgman growth process, sample S1, shows much lower Mn concentration in SLs, 4.4 ± 0.7 at.%, affirming that the crystallization in temperature gradient leads to significant improvement of the $MnBi_2Te_4$ structure. Finally, samples S3 and S4 show 5.0 ± 0.8 at.% and 7.6 ± 1.1 at.% of Mn in SLs, respectively (Table 1). Remarkably, Mn is not only present in the $MnBi_2Te_4$ septuple layers, but also substitutes on Bi sites in the $Bi_2Te_3$ quintuple layers. The sparse Mn population in QLs is clearly indicated by the TEM- EDX analysis between 0.3 and 1 at. % in all samples. Figure 1(d) and (e) shows EDX map in high resolution compared to the corresponding high-angle annular dark field scanning transmission electron microscopy (HAADF-STEM) image showing Mn substituting Bi in a $Bi_2Te_3$ QL. Secondary ion mass spectroscopy (SIMS) data are further consistent with the EDX analysis. The SIMS depth profile of Mn and Bi, Figure 1 (g), reveals a fragment of the superlattice structure originating from groups of three SLs split by four QLs, which are further separated by larger QL blocks. The increase of the Mn signal intensity and the accompanying decline of the Bi intensity are clearly revealed. In some cases, peaks marking the position of a single SL are resolved as well, indicating highly ordered layered structure extending on an area covered by the beam spot of one squared mm in diameter. The intensity of the Mn SIMS signal between peaks (in QL regions) is about ten times lower than at the peaks, consistently with EDX evaluations of Mn concentration in SLs and in QLs, respectively. Mn preferentially substitutes on Bi sites in QLs which, however, does not change the general arrangement of the crystallographic structure. The ubiquitous Mn/Bi intermixing in the MnBi$_2$Te$_4$ family has been earlier suggested in the literature reports as well.[9,21,27] It will be discussed below that the disorder in the distribution of Mn, in particular doping of Mn into QLs and depletion of SLs affects strongly both the magnetic properties and the surface electron band structure.

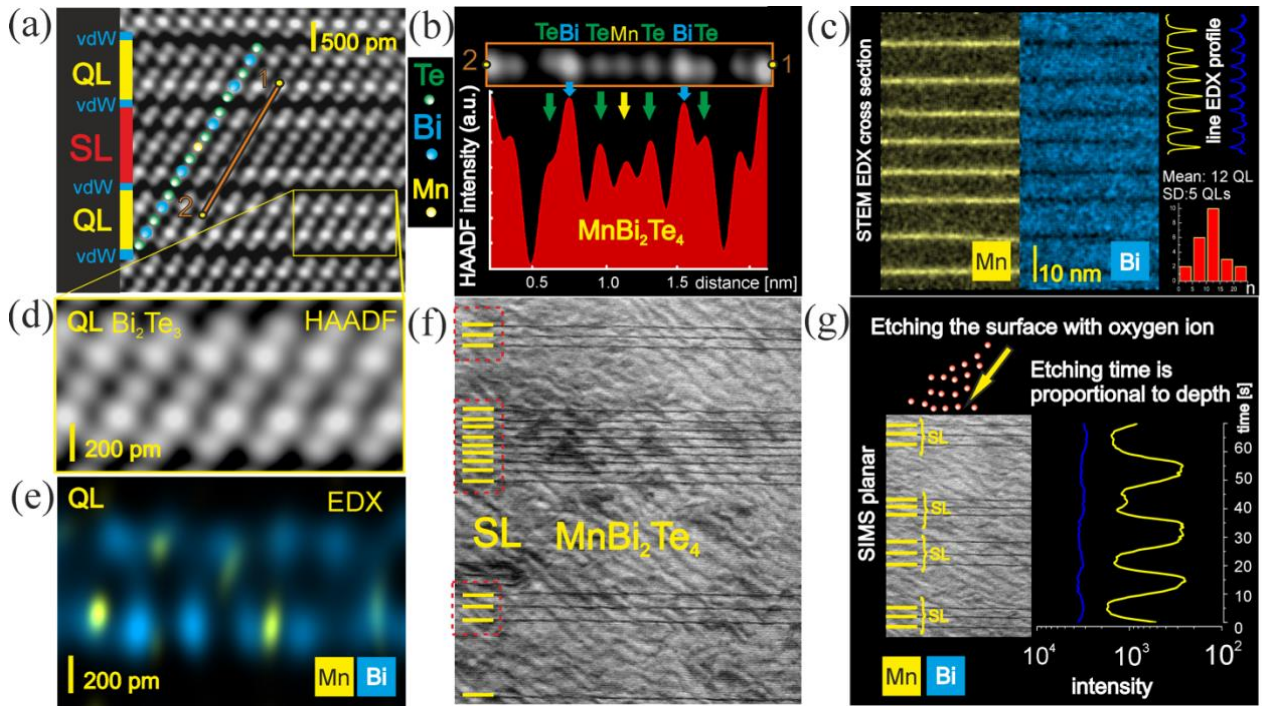

**Figure 1.** Structure and composition of MnBi$_2$Te$_4$/(Bi$_2$Te$_3$)$_n$. (a) HAADF-STEM image with intensity profile (b) of a SL in QL matrix. Bi as the heaviest element gives the highest intensity to the image. Manganese is incorporated in the middle layer of the SL. SL and neighboring QL are separated by the van der Waals gap. (c) EDX mapping in sample S1 with the distribution of Mn atoms shown in yellow and Bi atoms shown in blue, respectively. A histogram showing statistical distribution of the distance between SLs with mean distance 12 QLs and standard deviation 5 QLs is presented. (d) HAADF-STEM image and corresponding EDX mapping (e) showing Mn atoms substituting Bi site in a QL of sample S1. (f) STEM image showing grouping of SLs in superlattices containing multiple of 3 SLs in sample S2. The distance between SLs remains roughly constant, 4 QLs. (g) SIMS depth profile of sample S2 showing structures due to the superlattice formed by the SLs in the QL matrix. In QLs the concentration of Mn is about ten times lower than in SLs, consistently with EDX evaluations.

## 3. Disorder-Induced Magnetic Behaviors Detected in Ferromagnetic Resonance, FMR

It has been established that the intra-layer ordering within a model MnBi$_2$Te$_4$ single layer is ferromagnetic with phase transition temperature about 12 K, regardless the thickness of the MnBi$_2$Te$_4$ film or interfacing with other layered material through van der Waals gap.[10] In the three-dimensional MnBi$_2$Te$_4$ ($n$ = 0) built of SLs separated by van der Waals gaps, the inter-layer coupling is antiferromagnetic with bulk critical temperature increased up to 24 – 25 K by Anderson superexchange, which stabilizes the system and enhances the critical temperature above that for a single layer.[8,10] Weakness of the inter-layer coupling and its oscillating character in MnBi$_2$Te$_4$ ($n > 0$) does not allow to predict theoretically the FM or AFM order type and a critical temperature of the bulk magnetic phase for $n$ higher than 0, thus all the conclusions must rely on the experiment. In experiment, MnBi$_2$Te$_4$/(Bi$_2$Te$_3$)$_n$ with $n = 1 – 3$ show effects of inter-layer decoupling manifested by a strong drop of Néel temperature, $T_N$, with increased distance between SLs, from 25 K for $n = 0$ down to 13 K for $n = 1$ and 11.9 K for $n = 2$.[8] Remarkably, for $n = 2$ the FM bulk ordering has been reported next to the AFM,[17,27] which is consistent with similarly calculated energy of FM and AFM ground states,[8,28] making the system particularly vulnerable to disorder effects. For $n = 3$ and higher the vanishing exchange coupling between SLs has been reported.[8,28] The possibility of weak coupling of SLs via long-range RKKY interaction mediated by free carriers has not been excluded,[8] however, our earlier studies show that there is no evidence for that.[3] On the other hand, the role of Mn in QLs in stabilization of the bulk ferrimagnetic phase has recently been postulated.[17] This issue will be explored below.

In contrast to the AFM materials studied extensively before, samples investigated in this work, with $n$ between 2 and 12, show ferromagnetic response in magnetometry and/or magnetotransport with Curie temperatures between 6 K and 13 K, respectively (Table 1). The $T_c$ below the critical temperature of a single MnBi$_2$Te$_4$ layer can be accounted for missing Mn in a SL, as it is seen in structural studies (Sec. II), since the intra-layer exchange coupling will be reduced with highly increased distance between Mn sites. On the other hand, the presence of Mn in QLs cannot be ignored and the resulting critical temperature is thus a combined effect of the magnetic disorder in a single SL and the coupling of SLs via Mn in QLs. It has been established earlier that the pure QL-material Bi$_{2-x}$Mn$_x$Te$_3$ with Mn concentration around 0.8 – 1.8 at.% creates a ferromagnetic phase as well, with $T_c = 9 – 12$ K, respectively.[29,30]

It has been established, that magnetic resonance in two coupled ultra-thin films consists of two eigenmodes formed by the uniform modes of the individual layers, the acoustic mode for which the magnetization precession of both films occurs in-phase and the weaker optical mode where the mutual precession is out-of-phase.[23–26] In a case of ferromagnetically coupled films, the acoustic mode appears at higher magnetic field than the optical mode while for antiferromagnetic coupling the situation is opposite and the acoustic mode is located at lower magnetic field than the optical mode. This feature allows to distinguish the type of the coupling between magnetic layers.

The acoustic and optical modes of the coupled SL and Mn doped QL spin subsystems are clearly resolved at low temperatures for sample S2 ($n = 4$), which has the highest crystal uniformity followed by the lowest FMR line width (Figure 2 (a)). The stronger acoustic mode is located at higher magnetic field than the weaker optical mode allowing to assign the resonance to the ferromagnetic coupling. This finding contrasts with the recently proposed conceptual Ising model suggesting AFM coupling between SLs and Mn in QLs.[17] Here, the discrepancy may arise from the simplicity of the theoretical approach applied to the complex disordered spin system, in particular disregarding intra-layer disorder in a SL. A small difference in the amplitude ($I_{ac}/I_{opt} = 1.8$) and in the position of the two modes, Figure 2 (d), indicates a rather small value of the inter-layer coupling parameter, which will be discussed below. At 5 K, the two component lines can be resolved only around $H$ perpendicular to the Bi$_2$Te$_3$ $c$ axis, further the lines merge and the resonance is not detected around $H \parallel c$ due to large magnetic anisotropy constant and an insufficient applied microwave energy of the X-band, Figure 2 (b). At 11 K, however, one can recognize a typical anisotropy pattern, Figure 2 (c), of two coupled spin subsystems with the easy axis anisotropy. Remarkably, the two modes never cross in the angular dependence, which is characteristic for FMR in coupled films and allows to distinguish them from the uncoupled layers.

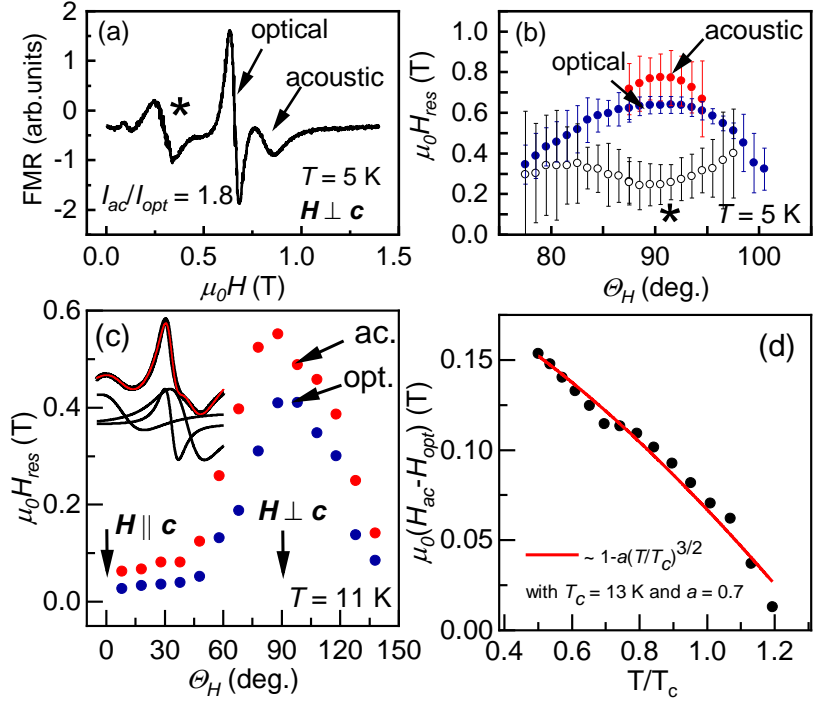

**Figure 2.** FMR in sample S2 showing separation of acoustic and optical resonance mode of coupled SLs and QLs. (a) FMR spectrum at 5 K around $H \perp c$ (90°). Optical and acoustic modes are indicated with arrows. Asterisk "*" marks extra resonance line originating from the second solution of the Smit and Beljers equations (see also Supporting Information and Figure S1). (b) Anisotropy of resonance signals at 5 K. Error bars indicate line width. (c) Anisotropy of the resonance signal at 11 K. The inset shows schematic decomposition of the measured signal into component lines. (d) Splitting between acoustic and optical modes at $H \perp c$ versus temperature.

## 4. Phenomenological Analysis of FMR in the Multilayered $MnBi_2Te_4/(Bi_2Te_3)_n$

According to a well-established phenomenological model for FMR in two exchange coupled thin films,[26] which can be adapted to multilayered material assuming that the spins in the respective subsystems ($MnBi_2Te_4$ and Mn doped $Bi_2Te_3$) see the same anisotropy and exchange fields,[31] respectively, for identical layers the higher lying acoustic mode is degenerate with that of a single layer while the optical mode is shifted from the single layer resonance towards a lower magnetic field by the exchange field $H_{ex}$ equal to (at $H \parallel ab$):

$$H_{ex} = \frac{2J}{M_s}\left(\frac{t_1+t_2}{t_1 t_2}\right). \quad (1)$$

Here, $J$ is the interlayer exchange energy per unit area, $M_s$ is the saturation magnetization, $t_1$ and $t_2$ are thicknesses of the respective magnetic layers. At 5 K, the observed splitting between the acoustic and the optical mode is about 150 mT, which assuming in rough approximation equal magnetic anisotropy constants for $MnBi_2Te_4$ and for Mn doped $Bi_2Te_3$, gives the estimation of the exchange coupling constant $J = 1.37 \times 10^{-7}$ J m$^{-2}$. Here, $t_1$ was assumed equal to the thickness of one SL, 1.68 nm, while $t_2$ to the thickness of four QLs, 5.12 nm, the saturation magnetization $M_s$ is 150 emu/mol.[3] The temperature variation of the splitting between the acoustic and the optical mode is shown in Figure 2 (d). The $T^{3/2}$ power dependence law can be applied, which according to [32] follows from thermal spin fluctuations at the interface that lead to the reduction of the effective interlayer coupling.

The splitting to the acoustic and the optical mode is clearly resolved in samples with higher $T_c$, while in lower $T_c$ samples the FMR appears as a single, irregular resonance line, Figure 3 (a), indicating weak coupling and similar magnetic anisotropy constants between $MnBi_2Te_4$ and Mn doped $Bi_2Te_3$. The FMR spectra can be conveniently approximated then by the uniform resonance mode with magnetic anisotropy constants treated as the magnetization-weighted mean values of the two spin subsystems.[25] The total free energy density $U$ which takes into account the dominant effect of the axial anisotropy along the $c$ axis can be expressed by:

$$U = -HM(\cos\theta\cos\theta_H + \sin\theta\sin\theta_H\cos(\varphi - \varphi_H)) - \frac{1}{2}M^2\sin^2\theta + K_1\sin^2\theta, \quad (2)$$

where the first term is the Zeeman term, the second term is the shape anisotropy energy and the third term is the axial anisotropy energy, approximated by the first element of a series expansion with anisotropy constant $K_1$. $M$ is sample magnetization. Angles $\theta$, $\varphi$, $\theta_H$, $\varphi_H$ are polar and azimuthal angles for $M$ and $H$ in the spherical coordinate system, respectively. The polar $z$ axis is along $Bi_2Te_3$ $c$ axis (the easy axis), the $xy$ plane lies in sample plane ($ab$ plane). Since demagnetization energy has the same dependence on $\theta$ as axial anisotropy energy, only the total effect is measured experimentally, which can be represented by the effective anisotropy field

$$H^A = \frac{K_1}{M} - \frac{1}{2}M. \quad (3)$$

In case of all studied $MnBi_2Te_4/(Bi_2Te_3)_n$, the easy axis is out-of-plane (along $c$ direction) due to large and positive $K_1$. Because the magnetocrystalline anisotropy dominates over the demagnetization anisotropy, the FMR occurs at lower magnetic field for $H$ applied parallel to the $c$ axis than for $H$ applied perpendicular to it, Figure 3 (a), like in $Bi_{2-x}Mn_xTe_3$.[29,30] This property caused great interest in this family of materials due to magnetic impact on the topological surface states caused by the out-of-plane component of magnetization.

The resonance field $H_{res}(\theta_H)$ can be then obtained by standard Smit and Beljers formula,[33]

$$\left(\frac{\omega}{\gamma}\right)^2 = \frac{1}{M^2\sin^2\theta}\left[\frac{\partial^2 U}{\partial\theta^2}\frac{\partial^2 U}{\partial\varphi^2} - \left(\frac{\partial^2 U}{\partial\theta\partial\varphi}\right)^2\right], \quad (4)$$

fulfilling simultaneously the conditions for minimum energy at equilibrium positions of magnetization, $\theta_{eq}$ and $\varphi_{eq}$,

$$\left(\frac{\partial U}{\partial\theta}\right)_{\theta=\theta_{eq},\varphi=\varphi_{eq}} = \left(\frac{\partial U}{\partial\varphi}\right)_{\theta=\theta_{eq},\varphi=\varphi_{eq}} = 0. \quad (5)$$

Here $\omega$ is the angular frequency of the applied microwaves and $\gamma$ is the gyromagnetic ratio. The resonance field varies with tilt angle $\theta_H$ around the average value $\omega\gamma^{-1}$, corresponding to $f = 9.5$ GHz of the applied microwave frequency and $g = 2.03$, extrapolated from the paramagnetic position of the resonance at higher temperatures, Figure 3 (b), which is a typical value for highly localized Mn center.[18] The equilibrium position of magnetization will be discussed in Supporting Information section. The difference between resonance field for $H$ applied parallel to the $c$ and perpendicular to it is three times the anisotropy field:

$$H_{res}(0°) - H_{res}(90°) = 3\,H^A. \quad (6)$$

The magnetic anisotropy field, $H^A$, determined from Equation 6 is shown in Figure 3 (c). The highest anisotropy field is obtained for sample S2, $\mu_0H^A = 280$ mT at 6 K, which according to Equation 3 corresponds to the anisotropy constant $K_1 = 407$ J m$^{-3}$ for $M$ being of the order of 150 emu/mol.[3] Typically, the anisotropy of the FMR signal is expected to scale with saturation magnetization.[34] However, while the magnetization tends to zero above the critical temperature in a way that only slightly deviates from mean-field-like manner,[3] the anisotropy of the FMR signal survives far above the $T_c$, up to about 40 K for all measured samples. This shows that the short-range magnetic correlations, with net-zero overall impact on sample magnetization, are present above the critical temperature. The effect may originate from the two-dimensional character of the studied system, since the two-dimensional ferromagnet in applied magnetic field can form large-scale correlations in the paramagnetic regime above the $T_c$, which are sensitive to the orientation of the applied field relative to the anisotropy axis.[35] These effects are typically observed in magnetic resonance of layered materials.[29,34,36]

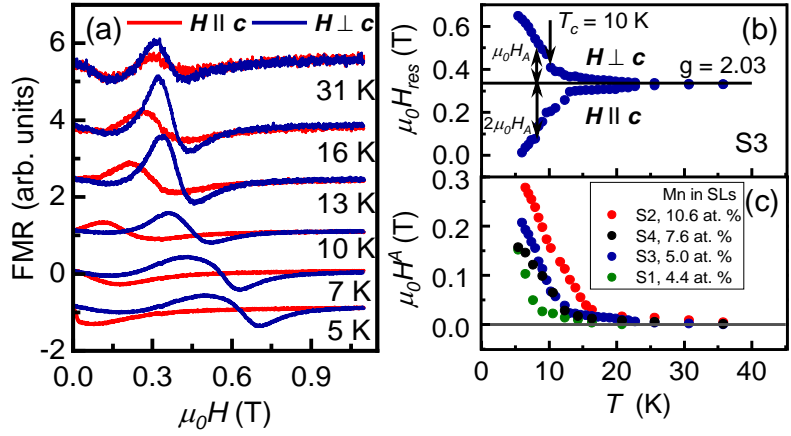

**Figure 3.** FMR of $MnBi_2Te_4/(Bi_2Te_3)_n$. (a) FMR spectra of sample S3 measured versus temperature for magnetic field applied parallel and perpendicular to the $Bi_2Te_3$ *c* axis. Single resonance line originates from similarity of magnetic anisotropy constants between $MnBi_2Te_4$ and Mn-doped $Bi_2Te_3$ and weak inter-layer coupling and allows to treat sample as a single-domain ferromagnetic material. (b) FMR anisotropy (resonance field for *H* || *c* and *H* ⊥ *c*, respectively) measured versus temperature for sample S3. (c) anisotropy field determined from Equation 6 versus temperature for S1 – S4 samples.

## 5. Surface Electronic Band Structure Under Disorder, ARPES and DFT

After establishing composition, structure and magnetic properties of $MnBi_2Te_4/(Bi_2Te_3)_n$, we studied the influence of the disorder on the surface states using ARPES measurements and DFT calculations. ARPES data obtained at photon energies from 8 eV up to 54 eV are shown in Figure 4 (a) – (f). Surface bands of the $MnBi_2Te_4/(Bi_2Te_3)_n$ family have complex orbital character which can be highlighted by appropriately selecting the sampling photon energy.[37] This is indeed visible in our data where intensities of the respective bands or their parts strongly vary with the applied photon energy. At 8 eV, the Dirac cone (TSS1) can be viewed analogous to topological surface states (TSSs) of pristine $Bi_2Te_3$,[38–40] with the Dirac point located about 0.35 eV below the Fermi level (Figure 4 (a)). However, applied higher photon energies (between 24 eV and 54 eV) reveal more complex structure in this region. In addition to the TSS1 states, a second Dirac cone (TSS2) and an extra parabolic band (CB1) appear, Figure 4 (e) and (f). Such a parabolic band has previously been reported in the literature[8,37,41] with an origin still under debate. Since DFT calculations of perfect structures do not reproduce the parabolic band, its origin was suggested to be a deviation from the ideal structure.[8,37,41] Finally, at intermediate photon energies (20 eV) a conduction band with linear dispersion (CB2) is observed. The measurement was done at about 6 K and all the bands remained essentially unchanged up to 120 K.

It has been already discussed in the literature[8,37] that surface states of $MnBi_2Te_4/(Bi_2Te_3)_n$ depend crucially on the surface termination type. One can distinguish a SL-terminated surface from a QL-SL-terminated surface. In the former, TSSs form a large magnetic gap of the order of 70 – 80 meV,[3,7,9,42] while in the latter, TSSs are hybridized with the valence band such that an anti-crossing gap is formed, which can be easily confused with the Dirac mass gap.[8,37] Finally for a QL-QL-SL-terminated surface, the surface bands resemble those for an unperturbed $Bi_2Te_3$.[8] All these features are clearly visible in the DFT calculations in Figure 4 (g) – (i). In Figure 4 (g), a symmetric SL-4QL-SL structure was calculated to show the effect of the opening of the magnetic gap on the SL-terminated surface. The accepted thickness of the slab is sufficient to simulate topological electronic structure.[43] Red color denotes surface states localized on the topmost SL while blue color denotes the states localized on the bottommost SL. A Dirac mass gap of about 70 meV is shown. Further, asymmetric slabs were investigated in order to assess disordered SL distribution in real crystals. QL-SL-2QL-SL and 2QL-SL-2QL-SL sequences were calculated as shown in Figure 4 (h) and (i), respectively. In these cases, the red and blue colors denote the states localized at the topmost QL and at the bottommost SL, respectively, while the orange color indicates states localized at the middle SL. This approach simultaneously simulates two different surface terminations. The anti-crossing of the Dirac cone (red) with valence band states (orange) characteristic for the QL-SL-terminated surface can be seen in Figure 4 (h). In Figure 4 (i), both the magnetic gap formed by blue bands of SL-QL-terminated surface and almost unperturbed $Bi_2Te_3$ Dirac cone (red) of 2QL-SL-terminated surface is viewed.

Remarkably, the electronic states of individual bands are all well localized on the respective QLs or SLs, in 40% or higher. The situation changes dramatically when disorder is introduced. In Figure 4 (j) 2QL-SL-2QL-SL structure was calculated with 50% of Mn atoms in SLs replaced by Bi. A first striking difference comparing to analogous perfect structure shown in Figure 4 (i) is delocalization of the wave functions. It is no longer possible to assign the respective bands to the corresponding QLs or SLs. Figure 4 (j) shows thus the states localized at the topmost QL (in more than 28%) in order to relate them to ARPES experiment probing finite depth of about 1 nm. The calculated band structure is qualitatively consistent with the ARPES data. One can identify the upper part of the parabolic band (CB1) while the bottom of the band has clearly a more bulk-like character. The double structure of the Dirac cone (TSS1 and TSS2) was also modeled successfully. Finally, the conduction band with linear dispersion (CB2) can be recognized as part of the Rashba-split band in Figure 4 (j). Furthermore, with disorder in SLs, the DFT band structure shows TSS1, TSS2, CB1, CB2 below the Fermi level (zero in Figure 4 (j)), which is consistent with the ARPES data.

Performed DFT calculations (not shown) confirm, that both FM and AFM coupling between SLs lead to analogous band structures. In contrary, neglecting magnetic coupling in a single SL causes the appearance of a gapless Dirac cone in the band gap on the SL-terminated surface. It means, that the disorder factor that can most strongly affect the structure of surface bands concerns the magnetic ordering within a single SL. Indeed, introducing magnetic disorder merely in a SL allowed to capture the most pronounced features in the ARPES data collected for the real world $MnBi_2Te_4/(Bi_2Te_3)_n$.

## 6. Discussion

The disorder changes systemically the properties of $MnBi_2Te_4/(Bi_2Te_3)_n$, affecting both the nature of the band structure through the delocalization of wave functions and introducing new bands, as well as influencing magnetic interactions — allowing the material to be treated as a stack of exchange coupled ultrathin layers. Quantitative description of the impact of the disorder on critical phase transition temperature is, however, challenging due to the number of factors that should be taken into account: the distance between SLs and its distribution, missing fraction of Mn in SLs and concentration of Mn in QLs as well as the thickness of the QL blocs (and its distribution). However, it is evident from Figure 3 (c) that the magnitude of the magnetic anisotropy field, tracked by the $T_c$, correlates with the quality of SLs. The highest $T_c$ = 13 K being obtained for sample S2 with evaluated concentration of Mn in SLs equal to about 10.6 at.%, while the lowest $T_c$ = 6 K for sample S1 with the highest deviation from the ideal structure, only 4.4 at.% of Mn in SLs. Sample S3 with 5 at. % of Mn in a SL is an intermediate case. In turn, sample S4 ($n = 2$) deserves particular attention since it has the highest overall concentration of Mn, 4 at.% (7.6 at. % in a SL), but does not show $T_c$ which is notably higher than its 2%-companion sample S3 (5 at.% of Mn in a SL). Here, evidently the shorter distance between SLs begins to play a role switching on AFM interactions between SLs with $n = 2$ and thus weakening FM exchange coupling. The obvious conclusion arises, that in order to engineer material with high ferromagnetic phase transition temperature, special attention should be paid to the quality of SLs while simultaneously performing doping of Mn into larger QL blocks.

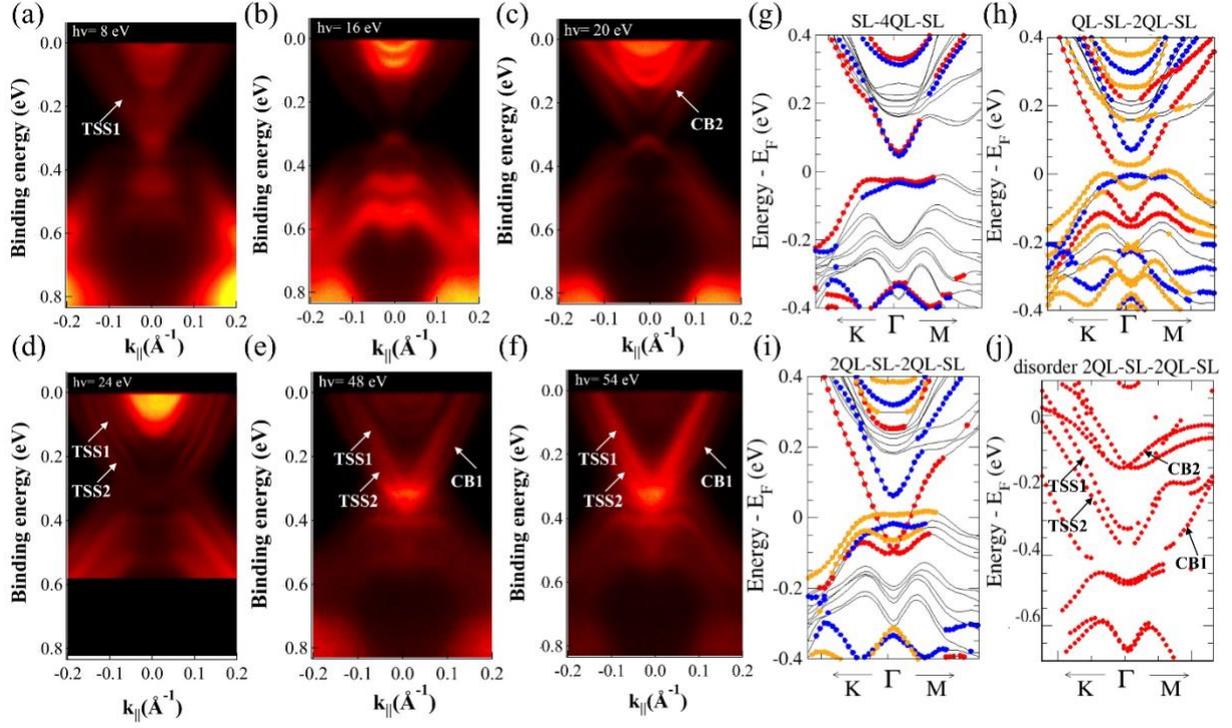

**Figure 4.** Surface electronic band structure of MnBi$_2$Te$_4$/(Bi$_2$Te$_3$)$_n$. (a)-(f) ARPES data of the disordered MnBi$_2$Te$_4$/(Bi$_2$Te$_3$)$_n$ (sample S3) obtained along $\bar{\Gamma} \to \bar{M}$ direction at photon energies 8 eV, 16 eV, 20 eV, 24 eV, 48 eV and 54 eV, respectively. (g)-(j) DFT-calculated band structures. The horizontal scale is the same as for ARPES. (g) SL(top)-4QL-SL(bottom), (h) QL(top)-SL-2QL-SL(bottom), (i) 2QL(top)-SL-2QL-SL(bottom), (j) 2QL-SL-2QL-SL with 50% of Mn atoms in SLs replaced by Bi. In (g) red color denotes surface states localized on the topmost SL while blue color denotes the states localized on the bottommost SL, (h)-(i), the red and blue color denote the states localized at the topmost QL at least in 40% and at the bottommost SL at least in 40%, respectively, while the orange color means states localized at the middle SL at least in 40%. In (j), only the states localized at the topmost QL in more than 28% are shown.

## 7. Conclusion

Summarizing, we analyzed structure, composition and magnetic properties of the ferromagnetic MnBi$_2$Te$_4$/(Bi$_2$Te$_3$)$_n$ with $n$ = 2, 4, 7 ± 1, and 12 ± 5. The complex system of the superlattice is strongly affected by disorder effects. In addition to the statistical distribution of the distance between SLs, which is the larger the higher is $n$, the omnipresent migration of Mn between SLs and QLs occurs: Mn is missing in SLs and substitutes into QLs. This has pronounced impact on magnetic properties as deviation from the ideal structure must lower the strength of intra-layer magnetic ordering within a single SL because of highly increased distance between planarly oriented Mn sites. Simultaneously the depleted SLs couple via an available population of Mn in QLs and the system is ferromagnetically stabilized in the three-dimensions. Disordered MnBi$_2$Te$_4$/(Bi$_2$Te$_3$)$_n$ typically appears as a single-domain ferromagnet, showing up in FMR as a uniform resonance mode, due to small inter-layer exchange coupling parameter (maximum value found $J = 1.37 \times 10^{-7}$ J m$^{-2}$) and similarity of magnetic anisotropy constants between MnBi$_2$Te$_4$ and Mn-doped Bi$_2$Te$_3$. Mn migration between QLs and SLs evidently makes MnBi$_2$Te$_4$/(Bi$_2$Te$_3$)$_n$ avoid magnetic differentiation in the component layers. We note that for samples with higher $T_c$ the optical and acoustic modes of two spin subsystems can be experimentally resolved.

The disorder changes the surface electronic structure, in particular causing appearance of a parabolic band CB1 and another trivial conduction band CB2 with linear dispersion, which can be easily confused with a gapped Dirac cone on a SL-terminated surface. Morover, the Dirac cone of QL-terminated surface evolves into peculiar double structure, which without a doubt requires further studies, in particular of its spin texture. Remarkably, all electron states that are well localized on respective QLs or SLs in perfect structures, in disordered material become delocalized. Finally, we remark that being aware and understand the impact of disorder on the surface states in intrinsic topological magnets is crucial to identify fingerprints for the large Dirac mass gap on the SL-terminated surfaces in ARPES experiments. The presence of such a gap is still under debate despite numerous studies performed.

## 8. Methods

*Sample Growth:* Samples for the studies were grown by vertical and horizontal variants of the Bridgman method in Kurnakov Institute of General and Inorganic Chemistry, Russian Academy of Sciences (vertical furnace) and in the Institute of Physics, Polish Academy of Sciences (horizontal furnace). The elements used in both growth processes were high purity (99.999 %) bismuth (Bi), tellurium (Te) and manganese (Mn). In the vertical method, the material was first synthesized in quartz ampules evacuated to $10^{-4}$ Pa. The ampules were loaded into a furnace, heated to 900 K, and maintained at this temperature for 48 h to achieve better homogenization. Then they were cooled down at a rate of 60 K per hour to 550 K and annealed for 24 hours. Then the furnace was switched off and cooled down to room temperature. In the next step, the synthesized material was ground and loaded into the evacuated ampules ($10^{-6}$ Torr) and sealed. To obtain a homogenized solution, the growing material was heated to 1073 K and rotated along the ampule axis for five days in the hot part of the furnace. Then the ampules were moved down at a speed of 2 mm per day. The temperature in the lower part of the furnace was kept at 873 K. This procedure resulted in crystals with average sizes of 50 mm in length and 14 mm in diameter. In the horizontal method, the quartz ampules sealed under vacuum ($10^{-6}$ Torr) were placed in the furnace containing two heating zones. In the first step, the ampules were heated up to a temperature of about 1053 K for 48 hours to synthesize the compound and to homogenize the melt. Then the ampules were cooled down to temperature of 903 K. Next, the ampules were pulled through the temperature gradient equal to 10 K per cm at a rate of 1 mm per hour. The single crystals obtained this way had average dimensions of 50 mm x 10 mm x 8 mm.

*Structural Studies:* Structural studies were performed using two magnification scales offered by scanning electron microscopy (SEM) and transmission electron microscopy (TEM). SEM investigations were performed using a Hitachi SU-70 microscope equipped with Thermo Scientific energy dispersive x-ray (EDX) spectrometer with silicon drift detector and Noran System 7 allowing for morphology and chemical composition studies. TEM investigations were carried out by the FEI Talos F200X microscope operated at 200 kV. High-resolution structural observations were performed in scanning transmission electron microscope (STEM) mode using a high-angle annular dark field (HAADF) imaging. EDX spectroscopy using a Super-X system with four SDDs was applied to detection of differences in local chemical composition. The samples for TEM investigations were cut along the *c*-axis, in the [11$\bar{2}$0] orientation, using a focused ion beam method. Additionally, secondary ion mass spectroscopy (SIMS) was carried out using a time-of-flight analyzer (IONTOF GmbH) enabling depth profiling.

*Ferromagnetic Resonance*: FMR measurements were performed using Bruker ELEXSYS-E580 electron paramagnetic resonance spectrometer operating in X-band (9.5 GHz), at temperatures varied by continuous-flow Oxford helium cryostat. Due to the use of magnetic field modulation and the lock-in technique the resonance signal represents field derivative of the absorbed microwave power.

*Angle-Resolved Photoemission Spectroscopy*: ARPES measurements were carried out at the National Synchrotron Radiation Centre SOLARIS in Cracow, Poland at the variable polarization UARPES beamline. Samples were glued with epoxy resin to a sample holder and cleaved in an ultra-high vacuum via mechanical exfoliation. As a photon source elliptically polarizing undulator (EPU) APPLE II type was used. Experimental data were collected by VGScienta DA30L electron spectrometer with an energy and angle resolution 3 meV and 0.1°, respectively.

*Density Functional Theory*: DFT calculations were performed using VASP code.[44,45] We used projector-augmented-wave pseudopotentials[46,47] and Perdew-Burke-Ernzerhof[48] generalized gradient approximation for the exchange-correlation functional. Spin-orbit coupling was included with an on-site Coulomb repulsion term U = 5.34 eV, chosen in order to take into account strong correlation of Mn d orbitals.[7] The structures of SL-4QL-SL and 2QL-SL-2QL-SL with and without Mn disorder were fully relaxed until the residual forces are less than 0.01 eV/Å, while the structure of QL-SL-2QL-SL was constructed from the experimental structures of $MnBi_2Te_4$ [49] and $Bi_2Te_3$.[50] We confirm that the effect of geometry relaxation is negligible. For the three structures without disorder, 11 x 11 x 1 k-points were sampled, while for the structure with 50% Mn disorder, 5 x 5 x 1 k-points were sampled since a supercell of 2x2 surface atoms was used to incorporate the disorder effect. In the supercell, 50% of the Mn atoms in each SL is replaced by Bi.

## Acknowledgements

This work was supported by the Polish National Science Center (NCN) grant 2016/21/B/ST3/02565. Computational support for Kyungwha Park was provided by Virginia Tech ARC and San Diego Supercomputer Center (SDSC) under DMR-060009N. Work at the CCNY was supported by NSF grants DMR-2011738 and HRD-1547830

## References

[1] M. Mogi, M. Kawamura, R. Yoshimi, A. Tsukazaki, Y. Kozuka, N. Shirakawa, K. S. Takahashi, M. Kawasaki, Y. Tokura, *Nat. Mater.* **2017**, *16*, 516.

[2] Ch.-X. Liu, Sh.-Ch. Zhang, and X.-L. Qi, *Annu. Rev. Condens. Matter Phys.* **2016**, *7*, 301.

[3] H. Deng, Z. Chen, A. Wołoś, M. Konczykowski, K. Sobczak, J. Sitnicka, I. V. Fedorchenko, J. Borysiuk, T. Heider, Ł. Pluciński, K. Park, A. B. Georgescu, J. Cano, L. Krusin-Elbaum, *Nat. Phys.* **2021**, *17*, 36.

[4] S. Grauer, K. M. Fijalkowski, S. Schreyeck, M. Winnerlein, K. Brunner, R. Thomale, C. Gould, L. W. Molenkamp, *Phys. Rev. Lett.* **2017**, *118*, 246801.

[5] D. Xiao, J. Jiang, J. H. Shin, W. Wang, F. Wang, Y. F. Zhao, C. Liu, W. Wu, M. H. W. Chan, N. Samarth, C. Z. Chang, *Phys. Rev. Lett.* **2018**, *120*, 56801.

[6] Y. Deng, Y. Yu, M. Z. Shi, Z. Guo, Z. Xu, J. Wang, X. H. Chen, Y. Zhang, *Science.* **2020**, *367*, 895.

[7] M. M. Otrokov, T. V. Menshchikova, M. G. Vergniory, I. P. Rusinov, A. Y. Vyazovskaya, Y. M. Koroteev, G. Bihlmayer, A. Ernst, P. M. Echenique, A. Arnau, E. V. Chulkov, *2D Mater.* **2017**, *4*, 025082.

[8] I. I. Klimovskikh, M. M. Otrokov, D. Estyunin, S. V Eremeev, S. O. Filnov, A. Koroleva, E. Shevchenko, V. Voroshnin, A. G. Rybkin, I. P. Rusinov, M. Blanco-rey, M. Hoffmann, Z. S. Aliev, M. B. Babanly, I. R. Amiraslanov, N. A. Abdullayev, V. N. Zverev, A. Kimura, O. E. Tereshchenko, K. A. Kokh, L. Petaccia, G. Di Santo, A. Ernst, P. M. Echenique, N. T. Mamedov, A. M. Shikin, E. V. Chulkov, *npj Quantum Mater.* **2020**, *5*, 54.

[9] M. M. Otrokov, I. I. Klimovskikh, H. Bentmann, D. Estyunin, A. Zeugner, Z. S. Aliev, S. Gaß, A. U. B. Wolter, A. V. Koroleva, A. M. Shikin, M. Blanco-Rey, M. Hoffmann, I. P. Rusinov, A. Y. Vyazovskaya, S. V. Eremeev, Y. M. Koroteev, V. M. Kuznetsov, F. Freyse, J. Sánchez-Barriga, I. R. Amiraslanov, M. B. Babanly, N. T. Mamedov, N. A. Abdullayev, V. N. Zverev, A. Alfonsov, V. Kataev, B. Büchner, E. F. Schwier, S. Kumar, A. Kimura, L. Petaccia, G. Di Santo, R. C. Vidal, S. Schatz, K. Kißner, M. Ünzelmann, C. H. Min, S. Moser, T. R. F. Peixoto, F. Reinert, A. Ernst, P. M. Echenique, A. Isaeva, E. V. Chulkov, *Nature* **2019**, *576*, 416.

[10] M. M. Otrokov, I. P. Rusinov, M. Blanco-Rey, M. Hoffmann, A. Y. Vyazovskaya, S. V. Eremeev, A. Ernst, P. M. Echenique, A. Arnau, E. V. Chulkov, *Phys. Rev. Lett.* **2019**, *122*, 107202.

[11] E. D. L. Rienks, S. Wimmer, J. Sánchez-Barriga, O. Caha, P. S. Mandal, J. Růžička, A. Ney, H. Steiner, V. V. Volobuev, H. Groiss, M. Albu, G. Kothleitner, J. Michalička, S. A. Khan, J. Minár, H. Ebert, G. Bauer, F. Freyse, A. Varykhalov, O. Rader, G. Springholz, *Nature* **2019**, *576*, 423.

[12] P. Wang, J. Ge, J. Li, Y. Liu, Y. Xu, J. Wang, *Innov.* **2021**, *2*, 100098.

[13] S. Wimmer, J. Sánchez-Barriga, P. Küppers, A. Ney, E. Schierle, F. Freyse, O. Caha, J. Michalička, M. Liebmann, D. Primetzhofer, M. Hoffmann, A. Ernst, M. M. Otrokov, G. Bihlmayer, E. Weschke, B. Lake, E. V. Chulkov, M. Morgenstern, G. Bauer, G. Springholz, O. Rader, (Preprint) arXiv:2011.07052, v2, submitted: Nov **2020**.

[14] N. De Jong, E. Frantzeskakis, B. Zwartsenberg, Y. K. Huang, D. Wu, P. Hlawenka, J. Sańchez-Barriga, A. Varykhalov, E. Van Heumen, M. S. Golden, *Phys. Rev. B* **2015**, *92*, 075127.

[15] J. Li, Y. Li, S. Du, Z. Wang, B. L. Gu, S. C. Zhang, K. He, W. Duan, Y. Xu, *Sci. Adv.* **2019**, *5*, eabg5685.

[16] S. V. Eremeev, M. M. Otrokov, E. V. Chulkov, *Nano Lett.* **2018**, *18*, 6521.

[17] C. Yan, Y. Zhu, S. Fernandez-Mulligan, E. Green, R. Mei, B. Yan, C. Liu, Z. Mao, S. Yang, (Preprint) arXiv:2107.08137, v1, submitted: Jul **2021**.

[18] A. Wolos, A. Drabinska, J. Borysiuk, K. Sobczak, M. Kaminska, A. Hruban, S. G. Strzelecka, A. Materna, M. Piersa, M. Romaniec, R. Diduszko, *J. Magn. Magn. Mater.* **2016**, *419*, 301.

[19] J. Růžička, O. Caha, V. Holý, H. Steiner, V. Volobuiev, A. Ney, G. Bauer, T. Duchoň, K. Veltruská, I. Khalakhan, V. Matolín, E. F. Schwier, H. Iwasawa, K. Shimada, G. Springholz, *New J. Phys.* **2015**, *17*, 9.

[20] I. Lee, C. K. Kim, J. Lee, S. J. L. Billinge, R. Zhong, J. A. Schneeloch, T. Liu, T. Valla, J. M. Tranquada, G. Gu, J. C. S. Davis, *Proc. Natl. Acad. Sci. U. S. A.* **2015**, *112*, 1316.

[21] Y. Liu, L. L. Wang, Q. Zheng, Z. Huang, X. Wang, M. Chi, Y. Wu, B. C. Chakoumakos, M. A. McGuire, B. C. Sales, W. Wu, J. Yan, *Phys. Rev. X* **2021**, *11*, 21033.

[22] C. Liu, Y. Ou, Y. Feng, G. Jiang, W. Wu, S. Li, Z. Cheng, K. He, X. Ma, Q. Xue, Y. Wang, *Phys. Rev. X* **2020**, *10*, 18.

[23] B. Heinrich, in *Ultrathin Magnetic Structures II* (Eds.: B. Heinrich, J. A. C. Bland), Springer-Verlag, Berlin, Heidelberg, **2005**, Ch. 3.

[24] J. Linder, K. Baberschke, *J. Phys. Condens. Matter* **2003**, *15*, S469.

[25] B. Heinrich, J. F. Cochran, *Adv. Phys.* **1998**, *42*, 523.

[26] Z. Zhang, L. Zhou, P. E. Wigen, K. Ounadjela, *Phys. Rev. B* **1994**, *50*, 6094.

[27] D. Souchay, M. Nentwig, D. Günther, S. Keilholz, J. De Boor, A. Zeugner, A. Isaeva, M. Ruck, A. U. B. Wolter, B. Büchner, O. Oeckler, *J. Mater. Chem. C* **2019**, *7*, 9939.

[28] J. Wu, F. Liu, C. Liu, Y. Wang, C. Li, Y. Lu, S. Matsuishi, H. Hosono, *Adv. Mater.* **2020**, *32*, 2001815.

[29] S. Zimmermann, F. Steckel, C. Hess, H. W. Ji, Y. S. Hor, R. J. Cava, B. Büchner, V. Kataev, *Phys. Rev. B* **2016**, *94*, 125205.

[30] Y. S. Hor, P. Roushan, H. Beidenkopf, J. Seo, D. Qu, J. G. Checkelsky, L. A. Wray, D. Hsieh, Y. Xia, S. Y. Xu, D. Qian, M. Z. Hasan, N. P. Ong, A. Yazdani, R. J. Cava, *Phys. Rev. B - Condens. Matter Mater. Phys.* **2010**, *81*, 195203.

[31] F. Keffer, C. Kittel, *Phys. Rev.* **1952**, *85*, 329.

[32] N. S. Almeida, D. L. Mills, M. Teitelman, *Phys. Rev. Lett.* **1995**, *75*, 733.

[33] J. Smit, H. G. Beljers, *Philips Res. Rep.* **1955**, *10*, 113.

[34] J. Zeisner, A. Alfonsov, S. Selter, S. Aswartham, M. P. Ghimire, M. Richter, J. van den Brink, B. Buchner, V. Kataev, *Phys. Rev. B* **2019**, *99*, 165109.

[35] M. G. Pini, P. Politi, R. L. Stamps, *Phys. Rev. B* **2005**, *72*, 014454.

[36] J. Zeisner, K. Mehlawat, A. Alfonsov, M. Roslova, T. Doert, A. Isaeva, B. Büchner, V. Kataev, *Phys. Rev. Mater.* **2020**, *4*, 64406.

[37] R. C. Vidal, H. Bentmann, J. I. Facio, T. Heider, P. Kagerer, C. I. Fornari, T. R. F. Peixoto, T. Figgemeier, S. Jung, C. Cacho, B. Büchner, J. Van Den Brink, C. M. Schneider, L. Plucinski, E. F. Schwier, K. Shimada, M. Richter, A. Isaeva, F. Reinert, *Phys. Rev. Lett.* **2021**, *126*, 176403.

[38] Y. L. Chen, J. G. Analytis, J. H. Chu, Z. K. Liu, S. K. Mo, X. L. Qi, H. J. Zhang, P. H. Lu, X. Dai, Z. Fang, S. C. Zhang, I. R. Fisher, Z. Hussain, Z. X. Shen, *Science.* **2009**, *325*, 178.

[39] C. Chen, S. He, H. Weng, W. Zhang, L. Zhao, H. Liu, X. Jia, D. Mou, S. Liu, J. He, Y. Peng, Y. Feng, Z. Xie, G. Liu, X. Dong, J. Zhang, X. Wang, Q. Peng, Z. Wang, S. Zhang, F. Yang, C. Chen, Z. Xu, X. Dai, Z. Fang, X. J. Zhou, *Proc. Natl. Acad. Sci. U. S. A.* **2012**, *109*, 3694.

[40] Y. Y. Li, G. Wang, X. G. Zhu, M. H. Liu, C. Ye, X. Chen, Y. Y. Wang, K. He, L. L. Wang, X. C. Ma, H. J. Zhang, X. Dai, Z. Fang, X. C. Xie, Y. Liu, X. L. Qi, J. F. Jia, S. C. Zhang, Q. K. Xue, *Adv. Mater.* **2010**, *22*, 4002.

[41] R. C. Vidal, H. Bentmann, T. R. F. Peixoto, A. Zeugner, S. Moser, C. H. Min, S. Schatz, K. Kißner, M. Ünzelmann, C. I. Fornari, H. B. Vasili, M. Valvidares, K. Sakamoto, D. Mondal, J. Fujii, I. Vobornik, S. Jung, C. Cacho, T. K. Kim, R. J. Koch, C. Jozwiak, A. Bostwick, J. D. Denlinger, E. Rotenberg, J. Buck, M. Hoesch, F. Diekmann, S. Rohlf, M. Kalläne, K. Rossnagel, M. M. Otrokov, E. V. Chulkov, M. Ruck, A. Isaeva, F. Reinert, *Phys. Rev. B* **2019**, *100*, 121104(R).

[42] J. Wu, F. Liu, M. Sasase, K. Ienaga, Y. Obata, R. Yukawa, K. Horiba, H. Kumigashira, S. Okuma, T. Inoshita, H. Hosono, *Sci. Adv.* **2019**, *5*, eaax9989.

[43] K. Park, J. J. Heremans, V. W. Scarola, D. Minic, *Phys. Rev. Lett.* **2010**, *105*, 186801.

[44] G. Kresse, J. Furthmüller, *Comput. Mater. Sci.* **1996**, *6*, 15.

[45] G. Kresse, J. Furthmuller, *Phys. Rev. B* **1996**, *54*, 11169.

[46] P. E. Blöchl, *Phys. Rev. B* **1994**, *50*, 17953.

[47] G. Kresse, D. Joubert, *Phys. Rev. B - Condens. Matter Mater. Phys.* **1999**, *59*, 1758.

[48] J. P. Perdew, K. Burke, M. Ernzerhof, *Phys. Rev. Lett.* **1996**, *77*, 3865.

[49] D. S. Lee, T. H. Kim, C. H. Park, C. Y. Chung, Y. S. Lim, W. S. Seo, H. H. Park, *CrystEngComm* **2013**, *15*, 5532.

[50] S. Nakajima, *J. Phys. Chem. Solids* **1963**, *24*, 479.

# Supporting Information

## Systemic Consequences of Disorder in Magnetically Self-Organized Topological MnBi$_2$Te$_4$/(Bi$_2$Te$_3$)$_n$ Superlattices

Joanna Sitnicka,[1] Kyungwha Park,[2] Paweł Skupiński,[3] Krzysztof Grasza,[3] Anna Reszka,[3] Kamil Sobczak,[4] Jolanta Borysiuk,[1] Zbigniew Adamus,[3] Mateusz Tokarczyk,[1] Andrei Avdonin,[3] Irina Fedorchenko,[5] Irina Abaloszewa,[3] Sylwia Turczyniak-Surdacka,[4] Natalia Olszowska,[6] Jacek Kołodziej,[6,7] Bogdan J. Kowalski,[3] Haiming Deng,[8] Marcin Konczykowski,[9] Lia Krusin-Elbaum,[8] and Agnieszka Wołoś[1*]

[1]Faculty of Physics, University of Warsaw, ul. Pasteura 5, 02-093 Warsaw, Poland

[2]Department of Physics, Virginia Tech, 850 West Campus Drive, Blacksburg, VA 24061, USA

[3]Institute of Physics, Polish Academy of Sciences, Aleja Lotników 32/46, PL-02668 Warsaw, Poland

[4]Faculty of Chemistry, Biological and Chemical Research Centre, University of Warsaw, ul. Zwirki i Wigury 101, 02-089 Warsaw, Poland

[5]Kurnakov Institute of General and Inorganic Chemistry, Russian Academy of Sciences, Leninskii prosp. 31, 117901 Moscow, Russia

[6]National Synchrotron Radiation Centre SOLARIS, Jagiellonian University, ul. Czerwone Maki 98, 30-392 Cracow, Poland

[7]Faculty of Physics, Astronomy and Applied Computer Science, Jagiellonian University, ul. prof. Stanisława Łojasiewicza 11, 30-348 Cracow, Poland

[8]Department of Physics, The City College of New York–CUNY, 85 St. Nicholas Terrace, New York, NY 10027, USA

[9]Laboratoire des Solides Irradiés, CEA/DRF/lRAMIS, Ecole Polytechnique, CNRS, Institut Polytechnique de Paris, F-91128 Palaiseau, France

*Corresponding author: agnieszka.wolos@fuw.edu.pl

## 1. Equilibrium Position of Magnetization

The magnetic anisotropy field, $H^A$, determines the preferred orientation of sample magnetization (in the uniform resonance mode approximation). For non-vanishing $H^A$, sample magnetization is aligned with the applied field only at $\theta_H = 0°$ ($H \parallel c$) and $\theta_H = 90°$ ($H$ in $ab$ plane), while at arbitrary angles the magnetization deviates from the applied field towards the easy axis and the effect is the larger the greater is the anisotropy field, Figure S1 (a). Remarkably, $\mu_0 H^A = 0.28$ T determined at 6 K for sample S2 is as large that the resonance condition cannot be fulfilled in the X-band for $H \parallel c$, while for $H$ in the $ab$ plane there occurs second solution of the Smit and Beljers equation corresponding to the second possible equilibrium angle, Figure S1 (a). The second resonance mode associated with the second possible equilibrium angle is marked with asterisk "*" in Figure 2 (a) and (b). To give impression of the impact of the magnetic anisotropy field on the direction of magnetization at constant magnetic field, Figure S1 (b) and (c) shows maps of the free energy density at magnetic field 0.3 T. Minimum of the energy determines equilibrium angle. For negligibly small $\mu_0 H^A = 0.01$ T the magnetization aligns with applied field while for $\mu_0 H^A = 0.28$ T the preferred easy axis along $c$ direction is evident. For $H$ in $ab$ plane there is actually no well-defined minimum of the free energy.

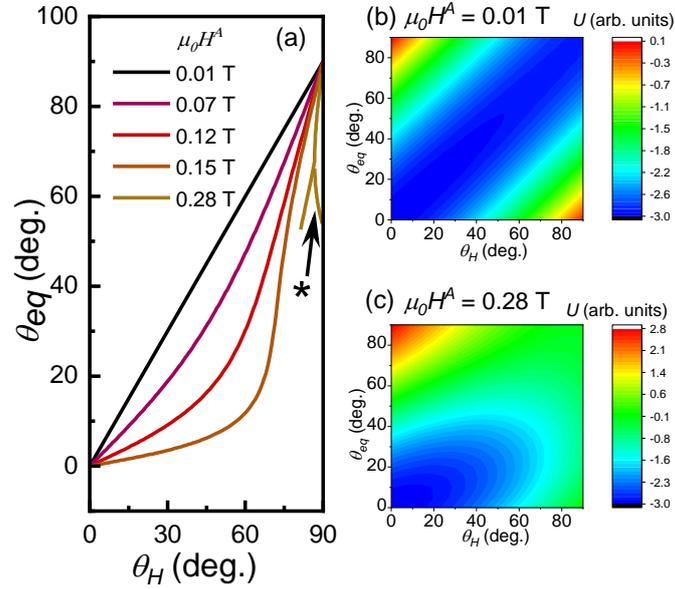

**Figure S1.** Impact of the magnetic anisotropy field on the equilibrium position of the magnetization. (a) Equilibrium angle of magnetization, $\theta_{eq}$, at the resonance field as a function of the angle between applied field and the $c$ axis, $\theta_H$, for assumed different anisotropy field $H^A$. (For anisotropy field 0.28 T two solutions of the Smit and Beljers problem are possible, lower branch marked with "*" is responsible for extra resonance line denoted in the same way in Figure 2 (a) and (b).) Free energy maps calculated for applied field 0.3 T for anisotropy field 0.01 T in (a) and 0.28 T in (b). Minimum of the energy indicates equilibrium position of magnetization.